%% file: root.tex
\documentclass[conference]{IEEE_ITSC}  

\IEEEoverridecommandlockouts                              

\input{packages}

\input{macros}

\definecolor{mn}{RGB}{255,127,0}
\definecolor{ts}{RGB}{0,0,255}
\definecolor{tosch}{RGB}{190,0,80}

\def\compileforpublish{0}

\renewcommand{\IEEEtitletopspaceextra}{20pt}
\title{Representing the Unknown -- Impact of Uncertainty on the Interaction between Decision Making and Trajectory Generation}

\author{Marcus Nolte, Susanne Ernst, Jan Richelmann and Markus Maurer\\
	
	\IEEEauthorblockA{Institute of Control Engineering\\
		Technische Universit\"at Braunschweig\\
		Braunschweig, Germany\\
		Email: \{nolte, ernst, richelmann, maurer\}@ifr.ing.tu-bs.de}
}

\begin{document}
\maketitle%
\thispagestyle{empty}%
\pagestyle{empty}%
\copyrightnotice%
\begin{abstract}%
\input{00_abstract}
\end{abstract}%
%
%
\section{Introduction}
\label{sec:intro}
\input{01_introduction}
\section{Related Work}
\label{sec:related}
\input{02_related_work}
\section{Challenges in Motion Planning}
\label{sec:challenges}
\input{03_challenges}

\section{Architectural Considerations}
\label{sec:architecture}
\input{04_architecture}
\section{Simulative Evaluation of an Exemplary MPC Planner}
\label{sec:results}
\input{05_framework}
\section{Conclusion \& Future Work}
\label{sec:conclusion}
\input{06_conclusion}
\section*{Acknowledgement}
\label{sec:ack}
\input{07_acknowledgement}

\IEEEtriggeratref{6}
\renewcommand*{\bibfont}{\footnotesize} 
\printbibliography 
\end{document}

%% file: packages.tex
\usepackage{tabularx,calc}
\usepackage{color}
\usepackage{tubscolors}

\usepackage{amsmath}
\usepackage{tikz}
\usetikzlibrary{shapes,automata,arrows,matrix,backgrounds,fit,patterns,decorations.markings, svg.path, shapes.multipart, external, shadows, positioning, calc, decorations}
\usepackage{pgfplots}
\usepackage{pgfplotstable}

\usepackage{siunitx}
\usepackage[style=ieee,backend=bibtex,minnames=1,maxcitenames=2,maxbibnames=99,doi=false,isbn=false,url=false,natbib=true]{biblatex} 
\usepackage[caption=false, font=footnotesize]{subfig}
\usepackage{ifthen}
\usepackage{lipsum}

\usepackage{amssymb} 	
\usepackage{bm} 		
\usepackage{booktabs} 	
\usepackage{multirow}   
\usepackage{colortbl}
\usepackage{array}
\usepackage{environ}

\AtBeginDocument{} 
\AtBeginDocument{} 

\newcolumntype{P}[1]{>{\centering\arraybackslash}p{#1}}

%% file: macros.tex
\makeatletter
\newcounter{IEEE@bibentries}
\renewcommand\IEEEtriggeratref[1]{%
	\renewbibmacro{finentry}{%
		\stepcounter{IEEE@bibentries}%
		\ifthenelse{\equal{\value{IEEE@bibentries}}{#1}}
		{\finentry\@IEEEtriggercmd}
		{\finentry}%
	}%
}
\makeatother

\ifx \isaccepted \undefined
\newcommand\copyrighttext{%
	\footnotesize \centering This work has been submitted to the IEEE for possible publication.\\ Copyright may be transferred without notice, after which this version may no longer be accessible.}
\else
\newcommand\copyrighttext{%
	\footnotesize \parbox[t]{.11\textwidth}{\copyright{} \the\year~IEEE.} \parbox[t]{.89\textwidth}{Personal use of this material is permitted. Permission from IEEE must be obtained for all other uses, in any current or future media, including reprinting/republishing this material for advertising or promotional purposes, creating new collective works, for resale or redistribution to servers or lists, or reuse of any copyrighted component of this work in other works.}}
\fi
\newcommand\copyrightnotice{%
	\ifx \compileforpublish \undefined
	\else
	\begin{tikzpicture}[remember picture,overlay]
	\node[anchor=south,yshift=10.5pt] at (current page.south) {\parbox{\dimexpr\textwidth-\fboxsep-\fboxrule\relax}{\copyrighttext}};
	\end{tikzpicture}%
	\fi
}

\makeatletter
\newsavebox{\measure@tikzpicture}
\NewEnviron{scaletikzpicturetowidth}[1]{%
	\def\tikz@width{#1}%
	\begin{lrbox}{\measure@tikzpicture}%
		\BODY
	\end{lrbox}%
	\pgfmathparse{#1/\wd\measure@tikzpicture}%
	\BODY
}
\makeatother

%% file: 00_abstract.tex
%
Even though motion planning for automated vehicles has been extensively discussed for more than two decades, it is still a highly active field of research with a variety of different approaches having been published in the recent years.
When considering the market introduction of SAE Level 3+ vehicles, the topic of motion planning will most likely be subject to even more detailed discussions between safety and user acceptance.
This paper shall discuss parameters of the motion planning problem and requirements to an environment model.
The focus is put on the representation of different types of uncertainty at the example of sensor occlusion, arguing the importance of a well-defined interface between decision making and trajectory generation.

%% file: 01_introduction.tex

Motion planning for automated vehicles is a topic which is currently receiving much attention in the scientific community.
A variety of frameworks has been presented in the past.
Those approaches range from discretized maneuver- or motion-primitive-based (e.g. \cite{werling2010, vonhundelshausen2008}) algorithms, to optimization-based approaches, such as optimal (receding horizon) control, often based on Model Predictive Control (MPC) \cite{andersen2017,gutjahr2016}.
Efforts are also spent on guaranteeing safety of motion planners in uncertain environments \cite{bouraine2012, magdici2016}.
Regarding the capabilities of the presented approaches, a majority of these planners is tailored to a specific application domain, such as parking or highway scenarios, while only some planners consider inner city scenarios.

One reason for these application specific frameworks is that motion planning is an optimization problem (cf.~\cite{benenson2008}).
However, the complete parameterization of this problem with respect to its cost-function and constraints is at least difficult, if not impossible without making simplifications. 
This leads to a variety of assumptions being made to simplify the optimization problem or to decompose it into solvable sub-problems which can be treated at different levels of the automation system.
Different assumption here lead to different motion planners.
The particular challenge for the parametrization of the optimization problem is the presence of uncertainty in the environment model \cite{benenson2008} (e.g.~measurement uncertainty and sensor occlusion), as well as in the motion dynamic objects (cf.~Section \ref{sec:architecture} for a detailed classification).

A central question with regard to the market introduction of automated vehicles and their user-acceptance must deal with how a motion planner treats this uncertainty.
A main challenge in this context will be the implementation of motion planners which are safe and not annoying for the vehicle's passengers at the same time.
While it is questionable that such a motion planner is even implementable, it becomes important to discuss the parameters of motion planning as well as the risks human drivers take and accept when entering public traffic.
An exemplary situation comprises driving at velocities which make it impossible to brake in front of movable objects emerging from occlusions.
While taking such risks when driving manually, people expect automated vehicles to be safe and to outperform human driving performance \cite{schoettle2014}.
Thus a very conservative (and thus probably safer) motion planner which calculates trajectories at slow speeds and stops the vehicle often will likely receive less acceptance than a planner which plans trajectories at higher velocities, risking collisions.

This is a twofold problem: In order to resolve those ambiguities in public perception, technical solutions are only partly sufficient.
In addition, expectation management for automated vehicle systems must be addressed in public discussions.
On the other hand, when automated vehicles are involved in (possibly deadly) accidents, the vehicle's driving strategy and behavior must be explained to the public and the authorities.
Those discussions will be difficult to lead at a mathematical level, such that semantic explanations of the executed decisions will be required, which can only be given at a tactical level of the system architecture.

In this paper we discuss the challenges for motion planning arising from an uncertain and incomplete environment representation, extending the argumentation of \cite{benenson2008} and \cite{shalev-shwartz2017}.
Possibilities to exploit an explicit representation of uncertainty are presented at different levels of a system architecture at the example of sensor occlusion.
In this context, we highlight the importance of a clear separation between decision making and trajectory generation in order to explain the generated behavior of an automated vehicle.

For this purpose, Section \ref{sec:related} discusses related work with respect to the formulation of the motion planning problem.
Section \ref{sec:challenges} presents the challenges when dealing with uncertainty and specifically sensor occlusion in more detail.
Based on this, Section \ref{sec:architecture} discusses these challenges from a system's perspective at different architectural levels, before Section \ref{sec:results} presents a possibility to incorporate the architectural considerations in an MPC framework.
Section \ref{sec:conclusion} concludes the paper with a summary and future work.

%% file: 02_related_work.tex
\citeauthor{benenson2008} formally define motion planning as a constraint optimization problem, while having a strong focus on motion safety \cite{benenson2008}.
In this context, formal safety criteria are given.
Additionally, they give a definition of an \emph{incomplete} and \emph{uncertain world model}, which is the consequence of occlusions, a limited sensor range and measurement uncertainties, respectively.
For a \emph{conservative world model}, probability distributions representing the uncertain world model are thresholded.
The authors take the approach of formulating a cost term for all safety-related assumptions.

\citeauthor{thornton2017a} present a motion planning framework based on Model Predictive Control \cite{thornton2017a}.
In contrast to \cite{benenson2008}, constraints are explicitly formulated as hard and soft constraints.
This comprises the need for collision-free trajectories (hard constraints) as well as ethical considerations and traffic rules (soft constraints).

This argumentation is picked up in \cite{bouraine2012}.
Based on the conservative world model, a notion of provably safe motion planning for robots with limited sensor range is developed.
The paper shows how the planning space in which safety guarantees can be given shrinks in a dynamic environment when calculating the reachable sets of moving obstacles with conservative models.

A technique to deal with this issue in the context of automated driving is to apply maneuver-based motion primitives and assume that other vehicles adhere the given traffic rules (e.g. as proposed in \cite{magdici2016, benenson2008}). 
However, it can be argued that an automated vehicle must also be able to react to non rule-conforming behavior of other vehicles, which then requires the fully conservative prediction strategy \cite{shalev-shwartz2017}.

\citeauthor{shalev-shwartz2017} propose a formal framework for formulating common sense and responsibilities in case of an accident \cite{shalev-shwartz2017}.
For this purpose, they formulate safety metrics in terms of distances and time-to-collision for longitudinal and lateral maneuvers.
While the core of their argumentation for formal verification is not reflecting realistic operating conditions, such as assuming that all sensors work correctly while trying to formalize safety and comfort aspects, they contribute some interesting metrics for increasing behavioral safety.
In addition to those metrics, the authors also deal with the problem of occluded space.
For this, they derive a maximal velocity at which occluded areas may be approached, considering that pedestrians suddenly emerge from an occlusion. 

Occlusions are also explicitly addressed in \cite{chung2009} in the context of path planning for an indoor robot.
Occluded areas are modeled in an occupancy grid.
A safety distance is calculated, assuming a maximum velocity for hidden moving obstacles.
The resulting area is avoided by introducing a cost term in a cost function for shortest-path-navigation.

\citeauthor{andersen2017} also highlight the importance of modeling occlusions, while presenting a Model Predictive Control framework minimizing the occluded area by performing lane changes \cite{andersen2017}.

While all of the previous approaches have contributed to the motion planning problem in some ways, the practical implication of the formulated issues from a system's perspective have rarely been discussed, if not completely neglected.

%% file: 03_challenges.tex
As discussed in \cite{benenson2008}, from a mathematical point of view, motion planning is an optimization problem.
Solving this optimization problem yields a nominal trajectory for the automated vehicle.
By controlling the vehicle to this desired trajectory, the system generates \emph{external behavior} \cite{nolte2017a}.
\emph{Unsafe external behavior}, or \emph{unsafe motion} can cause harm (as defined in \cite{benenson2008}).
For the description of safe motion, it is thus sufficient to only consider the system's trajectory $\boldsymbol{\pi}(t)$ as a series of system states $\mathbf x(t)$.

For a more general formulation of the motion planning problem as an optimal control problem, we consider a non-linear control system with inputs $\mathbf u(t)$ and outputs $\mathbf y(t)$ given by 
\begin{eqnarray}%
	\mathbf{\dot x}(t) =& \mathbf f(\mathbf x(t), \mathbf u(t)) \\
	\mathbf y(t) =& \mathbf g(\mathbf x(t), \mathbf u(t))
\end{eqnarray}

In the tuple notation of \cite{benenson2008}, the vehicle's trajectory is then formulated as
\begin{equation}
	\boldsymbol{\pi}(t) = \{(\mathbf g(\mathbf x(t_i), \mathbf u(t_i)), t_i)\, |\, i \in \mathbb{N}_0\}.
\end{equation}

Using the cost function $c(\boldsymbol \pi)$ and extending the notation of \cite{benenson2008}, we write the motion planning problem in a general form with time-varying hard equality and inequality or box constraints $\bm k_{\text{eq}}$ and $\bm k_{\text{ineq}}$
\begin{eqnarray}
	     \min\limits_{\mathbf u(t), \mathbf x(t)}{\,c(\bm \pi)}& \nonumber \\
	s.t. \qquad  \mathbf k_{\text{eq}}(t) =&  \mathbf k_{\text{constr}}(t), \\
		  \mathbf k_{\text{ineq}}^{\text{min}}(t) \leq& \mathbf k_{\text{ineq}}(t) \leq \mathbf k_{\text{ineq}}^{\text{max}}(t). \nonumber
\end{eqnarray}

This formulation is extended by slack variables $\bm \varepsilon(t)$ for including soft constraints, as e.g. proposed in \cite{thornton2017a}.
In this case the cost function is modified, weighting the slack variables for each soft constraint with a factor $s$ , to
\begin{equation}
	c'(\boldsymbol \pi)  = c(\boldsymbol{\pi}) + c_{\text{slack}}(\bm \varepsilon(t), \bm s),
\end{equation}
while adding constraints of the form
\begin{eqnarray}
	k_{\text{ineq},i}^{\text{min}} (t) - \varepsilon_i(t) \leq& k_{\text{ineq,i}} &\leq k_{\text{ineq},i}^{\text{max}} (t) + \varepsilon_i(t), \nonumber\\
	\varepsilon_i \geq& 0&
\end{eqnarray}
for the $i$-th added soft constraint.

The cost function can include a variety of requirements such as weights for passenger comfort (e.g. by weighting jerk), mission goals (e.g. shortest path to a target pose) and safety requirements (distance from other vehicles).
In cooperative systems, the cost function for the ego vehicle is also influenced by the goals and requirements of other traffic participants.

The challenge for the vehicle automation system now lies in a suitable parametrization of the constraints as well as of the weights in the cost function.
While this is obvious for motion planners which are based on continuous optimization methods, all other types of planners solve a simplified formulation of this problem, e.g. by using a-priori knowledge, such as a discretized state space (state-lattice and related approaches).

Regardless of the actual implementation, all motion planning algorithms must deal with uncertainty \cite{benenson2008}.
Considering the explainability of vehicle behavior, it is beneficial to differentiate two different types of uncertainty:

\emph{Epistemic} uncertainty \cite{kiureghian2009} describes missing information, which would be required to build a perfect world- and self-model.
For the world model, this uncertainty is based on non-observable parts of the environment caused by limited sensor range or occlusions, as well as missing knowledge about future actions of other traffic participants.
It is referred to as the \emph{incomplete world model} in \cite{benenson2008}.
Epistemic uncertainty also applies to the ego vehicle, e.g. in the form of unobservable system states or due to incomplete models arising from simplifications or errors when designing the vehicle model.
When planning motion, it is thus always important to ensure that a generated plan is executable by the automated vehicle.

\emph{Aleatoric} uncertainty, in contrast, is of statistical origin \cite{kiureghian2009}.
It thus e.g. includes measurement and process noise when modeling objects or the ego vehicle.
Regarding the environment model, it is referred to as the \emph{uncertain world model} in \cite{benenson2008}.

Epistemic uncertainty is often modeled in an aleatoric manner by introducing probability densities for unknown aspects of the environment (such as existence probabilities for perceived objects).
A clear separation between these types of uncertainty is thus debatable in some cases \cite{kiureghian2009}.
However, both categories can help making modeling decisions more explicit and thus making system behavior more explainable, as will be discussed below.

Both types of uncertainty make the parameterization of the motion planning problem difficult. 
Epistemic uncertainty is the reason for an inherently incomplete configuration of constraints, as not all possibly relevant objects are included in the world model.
At the same time aleatoric uncertainty allows only conservative assumptions about the evolution of the perceived environment.
In this context, a \emph{conservative world model} is proposed in \cite{benenson2008} by thresholding probability distributions for the environment.
However, the application of conservative predictions only enables safety guarantees for minimal parts of the planning space, as can be seen in \cite{bouraine2012}.
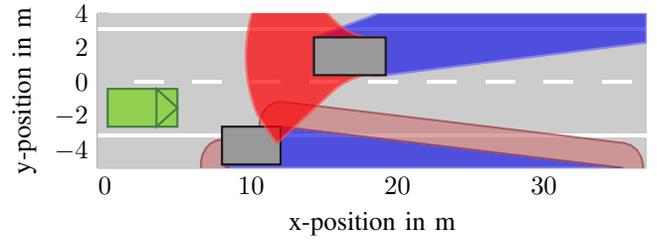
\begin{figure}
	\input{figures/tikz/overview}
	\caption{Scenario illustrating the consequences of epistemic and aleatoric uncertainty in the world model. The ego vehicle is depicted in green. Occlusions are marked blue, possible space occupied by the oncoming vehicle under aleatoric motion uncertainty in red, and possible occupation of unknown objects, such as pedestrians in light red. The resulting occupancies result from explicitly modeling epistemic relations in an aleatoric way. Prediction horizon: \SI{1}{\second}.} 
	\label{fig:simple_problem}
	\vspace{-1.5em}
\end{figure}

This already becomes obvious, when considering the following simple inner city scenario, as shown in Figure~\ref{fig:simple_problem}:
Given a two lane road with oncoming traffic, a van is parked at the right side of the road, occluding parts of the sidewalk.
A conservative system must account for objects such as pedestrians emerging from the occluded area in any direction at a wide range of possible velocities.

The occluded areas introduces epistemic uncertainty to the scene.
When introducing the occupancy of possible emerging objects, epistemic relations are modeled in an aleatoric fashion by making assumptions about possible accelerations and velocities of objects.
This in turn can then help when arguing, e.g. why an automated vehicle decelerates or performs a lane change for no immediately apparent reason.

Figure~\ref{fig:simple_problem} depicts the scenario with the predicted occupancies after one second.
A worst-case maximal acceleration $a_{p,\text{wc}}$ of \SI[per-mode=symbol]{0.3}{\meter\per\second^2} and an initial velocity of $v_{p,0}$ of \SI[per-mode=symbol]{1.3}{\meter\per\second} are assumed for potential objects in the occupied area.
For the vehicle we assume a maximal acceleration of \SI[per-mode=symbol]{0.1}{\meter\per\second^2} and an initial velocity of \SI[per-mode=symbol]{4.7}{\meter\per\second}, which is already a very conservative motion model.
The occupied area ends \SI[per-mode=symbol]{5}{\meter} in front of the ego vehicle.
For this illustration, no (traffic-)rule-based constraints are imposed on the vehicle's motion.
If we additionally assume a maximal deceleration of \SI[per-mode=symbol]{9}{\meter\per\second^2} for the ego vehicle, given dry asphalt as the road surface, at \SI[per-mode=symbol]{10}{\meter\per\second}, a safety guarantee cannot be given for the planned trajectory, as the stopping distance for the ego vehicle is \SI{5.56}{\meter}.
When requiring safety guarantees for a trajectory, the optimization problem is over-constraint in this scenario with no feasible solution available.
Even when reducing the maximal acceleration of the oncoming vehicle or increasing the possible deceleration for the ego vehicle, the only safe action in this scenario is stopping.

This example shows that even a very simple scenario, which is encountered regularly in public traffic, causes a dilemma for the development of motion planning algorithms between safety and mobility.
Even for conservative planning, safety cannot be fully guaranteed in every possible scenario.
As motion planning is an iterative problem, one could argue that a conservative approach will already react before coming into a situation where safety cannot be guaranteed.
However, such approaches will implement systems which drive slowly and stop often, which will impede user acceptance.
In addition, absolute safety can never be proven, as unexpected situations e.g. loss of cargo, trees falling onto the road (described as the \emph{falling crane scenario} in \cite{benenson2008}) can always occur.
Approaches, which apply traffic rules (e.g. as argued by \cite{benenson2008},\cite{althoff2016}) for the prediction of other traffic participants can only give weakened safety guarantees, but allow for higher velocities of the ego vehicle.

Vehicle-to-x communication could be seen as a means to limit aleatoric uncertainty (e.g. by including additional measurements in the world model) and epistemic uncertainty (by exchanging planned trajectories with other traffic participants).
However, one must keep in mind that information received from other entities can always be flawed and that high market penetration is required to fully exploit the benefits of V2X.
Thus it must never serve as the single input for safety-critical decisions and is thus also unsuitable when safety guarantees are required.

Looking at human drivers and the traffic system as a whole, it becomes clear, that human drivers always perform a personal trade-off between mobility and a need for safety. 
If human drivers showed a behavior as described above for a conservative planning approach, inner cities would be blocked by stopped vehicles.
Because of this fact, basic regulations such as the "principle of trust" or the requirement of "driving defensively" \cite{gasser2016} can be found in almost all international traffic laws in some form.
This relieves human drivers from expecting unreasonable behavior of other traffic participants and is thus a major contribution to a functioning traffic system.

While it can thus be argued that driving a vehicle seems to be an accepted risk, two key issues arise from these facts:
On the one hand, the question is what personal risk passengers are willing to accept when riding in an automated vehicle and if a principle of trust is applicable to a computer system.
On the other hand, it must be determined how a trade-off between a conservative and a more human-like motion planner can look like and which system-wide requirements arise for environment perception, scene representation and decision making in order to parameterize the motion planning problem accordingly.

One feasible approach can be the differentiation between a conservatively planned fall-back trajectory with (weakened) safety guarantees (cf. \cite{magdici2016}) in addition to a less conservatively planned trajectory.

Even with trajectory alternatives, a key requirement for planning both trajectories is the explicit representation of uncertainty.
While it is common to represent aleatoric uncertainty, e.g. as measurement noise in inverse sensor models for occupancy grids or model-based filtering algorithms for vehicle tracking, approaches which explicitly represent epistemic uncertainty (e.g. \cite{chung2009}, \cite{rieken2015a}, \cite{andersen2017}) are not as frequently applied in the field of automated driving.
For this paper, we assume a grid-based environment representation as presented in \cite{rieken2015a}, combining multiple grid layers such that drivable but occluded spaces can be identified as well as lane markings, curbs and higher elevated objects in the surroundings.
The fine grained semantic differentiation between elements in the grid representation is already an approach to represent and decrease epistemic uncertainty.

The following section will give an overview, how the suggested environment representation can be used for decision making and trajectory generation from an architectural point of view.
It will also argue for the importance of well-defined interfaces between decision making and trajectory generation.

%% file: figures/tikz/overview.tex
%
%
\definecolor{mycolor1}{rgb}{0.25098,0.50196,0.25098}%
\definecolor{mycolor2}{rgb}{0.57255,0.81569,0.31373}%
\definecolor{mycolor3}{rgb}{0.80000,0.50000,0.50000}%
\begin{tikzpicture}

\begin{axis}[%
width=2.93in*0.98,
height=0.704in*1.15,
at={(1.083in,1.381in)},
scale only axis,
xmin=-0.5,
xmax=37,
xlabel style={font=\color{white!15!black}},
xlabel={x-position in m},
ymin=-5,
ymax=4,
ylabel style={font=\color{white!15!black}},
ylabel={y-position in m},
axis background/.style={fill=white},
axis x line*=bottom,
axis y line*=left,
    ylabel near ticks,
xlabel near ticks,
]

\addplot[area legend, line width=1.5pt, draw=white!80!black, fill=white!80!black, forget plot,thin]
table[row sep=crcr] {%
x	y\\
-0.5	-5\\
37	-5\\
37	4\\
-0.5	4\\
}--cycle;

\addplot[area legend, draw=none, fill=white, forget plot]
table[row sep=crcr] {%
x	y\\
2	-0.1\\
2	0.1\\
4	0.1\\
4	-0.1\\
}--cycle;

\addplot[area legend, draw=none, fill=white, forget plot]
table[row sep=crcr] {%
x	y\\
6	-0.1\\
6	0.1\\
8	0.1\\
8	-0.1\\
}--cycle;

\addplot[area legend, draw=none, fill=white, forget plot]
table[row sep=crcr] {%
x	y\\
10	-0.1\\
10	0.1\\
12	0.1\\
12	-0.1\\
}--cycle;

\addplot[area legend, draw=none, fill=white, forget plot]
table[row sep=crcr] {%
x	y\\
14	-0.1\\
14	0.1\\
16	0.1\\
16	-0.1\\
}--cycle;

\addplot[area legend, draw=none, fill=white, forget plot]
table[row sep=crcr] {%
x	y\\
18	-0.1\\
18	0.1\\
20	0.1\\
20	-0.1\\
}--cycle;

\addplot[area legend, draw=none, fill=white, forget plot]
table[row sep=crcr] {%
x	y\\
22	-0.1\\
22	0.1\\
24	0.1\\
24	-0.1\\
}--cycle;

\addplot[area legend, draw=none, fill=white, forget plot, thick]
table[row sep=crcr] {%
x	y\\
26	-0.1\\
26	0.1\\
28	0.1\\
28	-0.1\\
}--cycle;

\addplot[area legend, draw=none, fill=white, forget plot, thick]
table[row sep=crcr] {%
x	y\\
30	-0.1\\
30	0.1\\
32	0.1\\
32	-0.1\\
}--cycle;

\addplot[area legend, draw=none, fill=white, forget plot, thick]
table[row sep=crcr] {%
x	y\\
34	-0.1\\
34	0.1\\
36	0.1\\
36	-0.1\\
}--cycle;

\addplot[area legend, line width=0.0pt, draw=none, fill=white, forget plot, thick]
table[row sep=crcr] {%
x	y\\
-0.5	-3\\
37	-3\\
37	-3.2\\
-0.5	-3.2\\
}--cycle;

\addplot[area legend, line width=0.0pt, draw=none, fill=white, forget plot, thick]
table[row sep=crcr] {%
x	y\\
-0.5	3\\
37	3\\
37	3.2\\
-0.5	3.2\\
}--cycle;

\addplot[area legend, line width=1.5pt, draw=mycolor1, fill=mycolor2, forget plot, thick]
table[row sep=crcr] {%
x	y\\
0.182	-2.6\\
0.182	-0.4\\
4.947	-0.4\\
4.947	-2.6\\
}--cycle;

\addplot[area legend, line width=1.5pt, draw=mycolor1, fill=mycolor2, forget plot, thick]
table[row sep=crcr] {%
x	y\\
3.5175	-0.4\\
4.947	-1.5\\
3.5175	-2.6\\
}--cycle;

\addplot[area legend, line width=1.5pt, draw=white!40!blue, fill=blue!80!gray, forget plot, thick, opacity=0.7]
table[row sep=crcr] {%
x	y\\
8.4061	-5\\
8	-4.8\\
8	-2.6\\
12	-2.6\\
35.3455	-5\\
}--cycle;

\addplot[area legend, line width=1.5pt, draw=white!40!blue, fill=blue!80!gray, forget plot, thick, , opacity=0.7]
table[row sep=crcr] {%
x	y\\
37	4\\
37	2.2919\\
19.188	0.4\\
14.283	0.4\\
14.283	2.6\\
18.7162	4\\
}--cycle;

\addplot[area legend, line width=1.5pt, draw=red!60!teal, fill=mycolor3, forget plot, thick, , opacity=0.7]
table[row sep=crcr] {%
x	y\\
8.4061	-6.45\\
8.3324039043917	-6.44812599089033\\
8.25889830141032	-6.4425088075753\\
8.18577319128979	-6.43316296957589\\
8.11321759075081	-6.4201126343894\\
8.04141904442286	-6.40339153504621\\
7.9705631400713	-6.38304289291527\\
7.90083302888263	-6.3591193059838\\
7.83240895204803	-6.33168261289985\\
7.76546777486859	-6.30080373312932\\
7.35936777486859	-6.10080373312932\\
7.00544447750029	-5.8551584301257\\
6.73931799451183	-5.51636644326652\\
6.5844812902475	-5.11433546147487\\
6.55460295240899	-4.68455575018571\\
6.65232056413955	-4.264967161607\\
6.86900786493925	-3.89260990173426\\
7.1855362548492	-3.6003547158285\\
7.57396341826592	-3.41400114320958\\
8	-3.35\\
8	-4.8\\
8.4061	-5\\
}--cycle;

\addplot[area legend, line width=1.5pt, draw=red!60!teal, fill=mycolor3, forget plot, thick, , opacity=0.7]
table[row sep=crcr] {%
x	y\\
10.55	-2.6\\
10.5749872326207	-2.33197273868853\\
10.6490877418016	-2.07318307672476\\
10.7697476424939	-1.83255023822366\\
10.9328083733092	-1.61836766968427\\
11.132650021955	-1.43801720512509\\
11.3623850163456	-1.29771465007879\\
11.614095505797	-1.20229555293191\\
11.8791062509256	-1.15504854703186\\
12.1482836170619	-1.15760200745043\\
35.4937836170619	-3.55760200745043\\
35.9632786686269	-3.68818845995717\\
36.3629953110419	-3.96694468105155\\
36.6477853499212	-4.3623850163456\\
36.7854816081446	-4.8298442824786\\
36.7605312245781	-5.31652272188415\\
36.5757523575062	-5.76744976177634\\
36.252015872246	-6.13169296779913\\
35.8258879664871	-6.36811088792333\\
35.3455	-6.45\\
35.3455	-5\\
12	-2.6\\
}--cycle;

\addplot[area legend, line width=1.5pt, draw=white!10!black, fill=white!60!black, forget plot, thick]
table[row sep=crcr] {%
x	y\\
8	-4.8\\
8	-2.6\\
12	-2.6\\
12	-4.8\\
}--cycle;

\addplot[area legend, line width=1.5pt, draw=white!30!red, fill=red, forget plot, thick, opacity=0.7]
table[row sep=crcr] {%
x	y\\
19.188	0.4\\
19.188	2.6\\
18.7556457726186	2.6334355713605\\
18.3178508303508	2.63793753927071\\
17.8818659936501	2.61288637514197\\
17.3434274005986	2.70506246977272\\
17.0467166077549	2.79726337956286\\
16.7557696934115	2.91060427721474\\
16.4716625146007	3.04547814838166\\
16.1954578030504	3.2023182926294\\
15.9282005606471	3.3815807454416\\
15.6709133894561	3.58373555168973\\
15.4074465763878	3.81468006009525\\
15.1439797633195	4.04562456850078\\
14.8805129502513	4.2765690769063\\
14.617046137183	4.50751358531182\\
14.3535793241147	4.73845809371734\\
14.0901125110465	4.96940260212286\\
13.8266456979782	5.20034711052838\\
13.5631788849099	5.43129161893391\\
13.2997120718416	5.66223612733943\\
13.0362452587734	5.89318063574495\\
12.7727784457051	6.12412514415047\\
12.5093116326368	6.35506965255599\\
12.2458448195685	6.58601416096151\\
11.9823780065003	6.81695866936704\\
11.7350766329746	6.55987144457667\\
11.4990137766634	6.29437332925807\\
11.2747803781774	6.02078154682955\\
11.1767272126516	5.8978536923025\\
11.0155970538818	5.71403269688959\\
10.8544668951121	5.53021170147668\\
10.6933367363424	5.34639070606377\\
10.5322065775726	5.16256971065087\\
10.372070080403	4.8329628952763\\
10.2267927433737	4.4995324888337\\
10.0967243476181	4.16277616770312\\
9.98217713366734	3.82320044310806\\
9.883424982344	3.48131931563674\\
9.80070269642639	3.13765290986894\\
9.73420538505248	2.79272609273527\\
9.68408795255307	2.44706707927897\\
9.65046469312236	2.10120602952401\\
9.64730968241707	1.90669313436841\\
9.63340899244732	1.75567364017649\\
9.6324132403144	1.56328351015825\\
9.63295313712755	1.41099973490052\\
9.63410983739908	1.2185271730173\\
9.65240529041124	0.872951885131717\\
9.68726423714797	0.527087584119787\\
9.73861022682101	0.181464975324145\\
9.80632591871453	-0.163385877935322\\
9.89025338858143	-0.506936962403133\\
9.99019454192449	-0.848663717484162\\
10.1059116330196	-1.18804642866804\\
10.2371278882517	-1.52457160808062\\
10.3835282320528	-1.85773335849518\\
10.5357114165553	-2.04902728096644\\
10.6878946010578	-2.2403212034377\\
10.8400777855602	-2.43161512590896\\
11.0783297265512	-2.76072067792689\\
11.2911161415499	-3.04150003557582\\
11.5162533848897	-3.31449841212898\\
11.7531768925751	-3.5793786607365\\
12.0273544967145	-3.3612576646371\\
12.3015321008539	-3.1431366685377\\
12.5757097049933	-2.92501567243829\\
12.8498873091327	-2.70689467633889\\
13.1240649132721	-2.48877368023949\\
13.3982425174115	-2.27065268414008\\
13.6724201215509	-2.05253168804068\\
13.9465977256903	-1.83441069194128\\
14.2207753298297	-1.61628969584187\\
14.4949529339691	-1.39816869974247\\
14.7691305381085	-1.18004770364306\\
15.0433081422479	-0.961926707543661\\
15.3174857463873	-0.743805711444258\\
15.5916633505267	-0.525684715344854\\
15.8597845569391	-0.336591141623017\\
16.1367468515878	-0.168680702042799\\
16.4216266791861	-0.0215599437882266\\
16.7134844152622	0.105151951320337\\
17.0113679925147	0.211815321829222\\
17.3143163722925	0.298753913643551\\
17.621361953642	0.3662206519328\\
17.8698440127632	0.387472943098005\\
18.3097091369891	0.359905367783965\\
18.7517228682214	0.364309411618514\\
19.188	0.4\\
}--cycle;

\addplot[area legend, line width=1.5pt, draw=white!10!black, fill=white!60!black, forget plot, thick]
table[row sep=crcr] {%
x	y\\
14.283	0.4\\
14.283	2.6\\
19.188	2.6\\
19.188	0.4\\
}--cycle;
\end{axis}
\draw[use as bounding box, white]  (1.6,5.75) rectangle (10.1,2.6);
\end{tikzpicture}%

%% file: 04_architecture.tex
%

As described in Section~\ref{sec:challenges}, motion planning is an optimization problem under the influence of uncertainty.
Design decisions about handling uncertainty in decision making and motion planning should be traceable.
Thus roles of modules should be explicitly defined in a functional system architecture, as the semantics of available information vary between upper and lower system layers (cf. \cite{nolte2017}).
However, this separation often remains unconsidered in recent publications (e.g. \cite{andersen2017}, \cite{thornton2017a}).

\begin{figure}[htbp]
	\centering
	\includegraphics[width=.98\columnwidth]{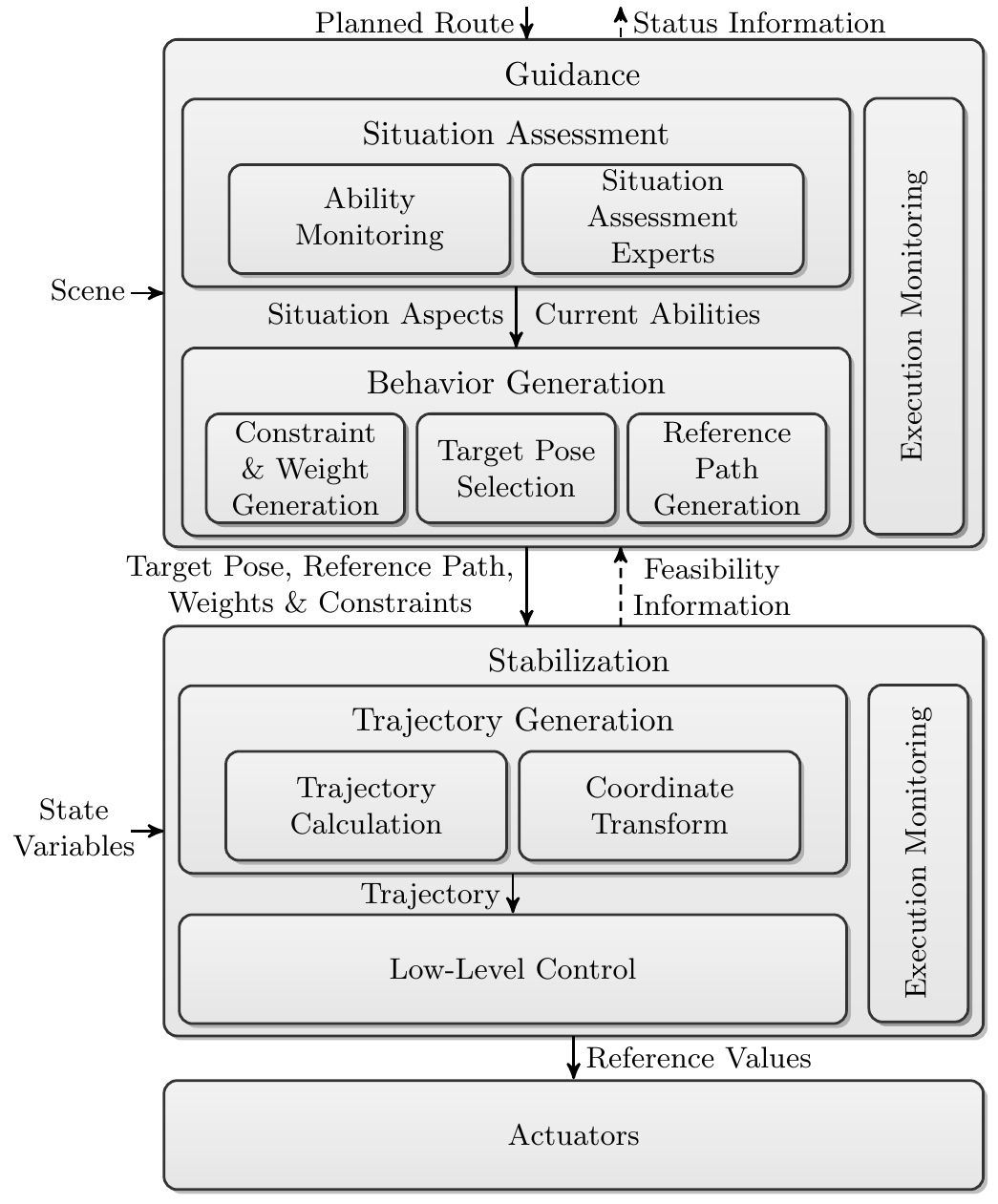}
	\caption{View on the tactical (guidance block) and stabilization level in a functional architecture with blocks contributing to decision making and motion planning (cf.~\cite{nolte2017}).}
	\label{fig:architecture_zoom}
	\vspace{-1.5em}
\end{figure}

Figure \ref{fig:architecture_zoom} shows logical components of the tactical and the stabilization level.
This follows the argumentation of \citeauthor{matthaei2015g} \cite{matthaei2015g} and \citeauthor{donges1999} \cite{donges1999} to organize the system according to how a human performs the driving task.
A detailed functional description of the tactical and the stabilization level can be found in \cite{nolte2017}.

The guidance block at the tactical level receives information about the current scene, which contains the environment model (including semantic relations) and information about the vehicle's current abilities \cite{Ulbrich2015a}, while the stabilization level only receives feature-level information.
This means that information, e.g. about lane markings, the relative position of relevant static and dynamic obstacles, and the relevant traffic rules for a given situation are only fully available at the tactical level.
In the following, we use an exemplary scenario in order to argue why this differentiation is required for traceable decision making.
We also highlight which consequences arise for the executed maneuvers and thus the external behavior of an automated vehicle.

For the examples at hand, we assume German traffic rules to apply.
Thus e.g. solid white lines may not be crossed without good reason and slower vehicles may not be passed on the right hand side.

\begin{figure}
	\centering
	\input{figures/tikz/target_pose_selection}
	\caption{Illustration of different target poses, which result in different maneuvers. The dotted target pose results in following the slower vehicle. The dashed target pose implies a lane change maneuver, as part of an overtaking maneuver. Skipping the dashed target pose results in an evasion maneuver.}
	\label{fig:maneuvers}
	\vspace{-1.5em}
\end{figure}
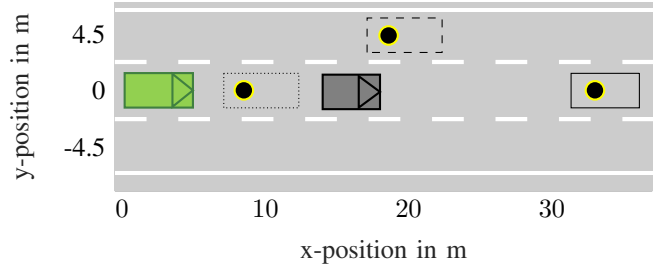

Consider Figure~\ref{fig:maneuvers} as an example with the ego vehicle approaching a slower vehicle on a three lane road.
In this example, the given situation can be resolved in two different ways: One option would be to stay behind the slower vehicle and perform a \emph{following} maneuver.
A different choice could be to perform a \emph{lane change} maneuver in order to overtake.

Both maneuvers can be achieved by at least two different strategies.
The system could choose a target pose at the end of the system's sensor range, assuming here that e.g. a sensor (e.g. a roof mounted LiDAR scanner) can perceive the space in front of the slower vehicle.
However, without careful parametrization of the algorithm, the executed maneuver becomes a result of the trajectory planner's cost-function, depending on which maneuver causes fewer optimization cost.
In the worst case, even the side on which the slower vehicle is passed, would remain undefined.
Considering differing traffic regulations in different countries, this is problematic, as passing on the right is e.g. prohibited in Germany, while being allowed in the U.S.

On the other hand, the system could explicitly choose a target pose behind the leading vehicle. 
This choice represents an explicit decision for \emph{following} the lead vehicle.
For performing an overtaking maneuver under European traffic rules, the target pose is placed in the left lane in front of the leading vehicle.
As soon as the left lane is reached, the target pose can be returned to the right lane, such that the ego vehicle can merge back. 

The latter way of executing the maneuver leaves the tactical layer direct control of the system's state.
When considering the first option, the choice of maneuvers is left to the stabilization level, which makes the system's behavior difficult to trace.

This differentiation becomes even more important if the decisions to make implies violating traffic rules.
Consider a lane bounded by a solid lane center line, which is blocked by a parked vehicle (cf.~\cite{thornton2017a}).
A trade-off must be made between crossing that line and stopping.
While in this specific example, legislations allow crossing solid lines by necessity, e.g. if a road would be impassible otherwise, the decision to violate a traffic rule must be an explicit decision, not a mere output of an optimization algorithm.
In turn, when resolving such situations only at the stabilization level, a non-carefully parametrized trajectory planner could plan a path which violates traffic rules without the tactical system modules being aware of this fact.
As a result, the system's tactical layer should have full control of the trajectory planner's cost-function and the relevant constraints, which derive from the chosen target pose and the semantics of the perceived environment.

If uncertainties are to be respected when making decisions, the described separation between tactical decisions and the bare execution of driving maneuvers becomes even more important.
In this case, frequent decisions must be made by weighting a conservative driving style against a more risk prone style.
Regrading the trajectory planner's cost function, the tactical modules must then be able to modify the cost terms to enable balancing between comfort and safety parameters (e.g. weighting of yaw-rates and side slip angle).
The control of hard constraints explicitly models rules which must be adhered to, such as collision avoidance requirements.
Soft constraints can be used to model rules which can be violated under circumstances by prioritizing constraints, e.g. when breaking a rule by necessity.

By explicitly parameterizing the optimization problem from the tactical level, the system can make active decisions which then reflect in the optimization problem.
If the system then decides to place the target pose at the end of the sensor range, objects in the ego lane can, of course, still be avoided.
However, the semantics of the executed maneuver changes: The result becomes an evasion rather than an overtaking maneuver.
While this example is arguably a minor distinction, making explicit decisions results in increasingly traceable system behavior.
Being able to trace and argue the resulting system behavior on both, a technical and a semantic level in case of errors can contribute to public discussions, when it becomes necessary to argue why an automated vehicle system behaved in a particular way.

%% file: figures/tikz/target_pose_selection.tex
%
%
\definecolor{mycolor1}{rgb}{0.80000,0.50000,0.50000}%
\definecolor{mycolor2}{rgb}{0.00000,0.44700,0.74100}%
\definecolor{mycolor3}{rgb}{1.00000,1.00000,0.00000}%
\definecolor{mycolor4}{rgb}{0.85000,0.32500,0.09800}%
\definecolor{mycolor5}{rgb}{0.25098,0.50196,0.25098}%
\definecolor{mycolor6}{rgb}{0.57255,0.81569,0.31373}%
\begin{tikzpicture}

\begin{axis}[%
width=2.93in*0.96,
height=0.704in*1.4,
at={(1.083in,1.381in)},
scale only axis,
xmin=-0.5,
xmax=37,
xlabel style={font=\color{white!15!black}},
xlabel={x-position in m},
ymin=-8,
ymax=4,
ylabel style={font=\color{white!15!black}},
ylabel={y-position in m},
ytick={-5.2, -1.6, 2},
yticklabels={-4.5,0, 4.5},
axis background/.style={fill=white},
axis x line*=bottom,
axis y line*=left
]

\addplot[area legend, line width=1.5pt, draw=white!80!black, fill=white!80!black, forget plot, thick]
table[row sep=crcr] {%
x	y\\
-0.5	-8\\
37	-8\\
37	4\\
-0.5	4\\
}--cycle;

\addplot[area legend, line width=0.0pt, draw=none, fill=white, forget plot, thin]
table[row sep=crcr] {%
x	y\\
-0.5	-6.8\\
37	-6.8\\
37	-7\\
-0.5	-7\\
}--cycle;

\addplot[area legend, line width=0.0pt, draw=none, fill=white, forget plot, thin]
table[row sep=crcr] {%
x	y\\
-0.5	3.4\\
37	3.4\\
37	3.6\\
-0.5	3.6\\
}--cycle;

\addplot [color=mycolor3, draw=none, mark size=3.5pt, mark=*, mark options={solid, fill=black}, forget plot, , thick]
  table[row sep=crcr]{%
33	-1.6\\
};

\addplot [color=mycolor3, draw=none, mark size=3.5pt, mark=*, mark options={solid, fill=black}, forget plot, , thick]
  table[row sep=crcr]{%
8.5	-1.6\\
};

\addplot [color=mycolor3, draw=none, mark size=3.5pt, mark=*, mark options={solid, fill=black}, forget plot, , thick]
  table[row sep=crcr]{%
18.6 	1.9\\
};

\addplot [color=black, forget plot, thin]
  table[row sep=crcr]{%
36.0988136223526	-0.515325813281867\\
36.0991862730082	-2.71532578172084\\
31.3341863413665	-2.71613290916368\\
31.3338136907109	-0.516132940724706\\
36.0988136223526	-0.515325813281867\\
};

\addplot [color=black, dashed, forget plot, thin]
  table[row sep=crcr]{%
17.0988136223526	0.815325813281867\\
17.0991862730082	3.01532578172084\\
22.3341863413665	3.01613290916368\\
22.3338136907109	0.816132940724706\\
17.0988136223526	0.815325813281867\\
};

\addplot [color=black, densely dotted, forget plot]
  table[row sep=crcr]{%
7.0988136223526	-0.515325813281867\\
7.0991862730082	-2.71532578172084\\
12.3341863413665	  -2.7153290916368\\
12.3338136907109	-0.516132940724706\\
7.0988136223526	-0.515325813281867\\
};

\addplot[area legend, line width=1.5pt, draw=mycolor5, fill=mycolor6, forget plot, thick]
table[row sep=crcr] {%
x	y\\
0.182	-2.7\\
0.182	-0.5\\
4.947	-0.5\\
4.947	-2.7\\
}--cycle;

\addplot[area legend, line width=1.5pt, draw=mycolor5, fill=mycolor6, forget plot, thick]
table[row sep=crcr] {%
x	y\\
3.5175	-0.55\\
4.947	-1.6\\
3.5175	-2.65\\
}--cycle;
\addplot[area legend, line width=1.5pt, draw=black, fill=gray, forget plot, thick, ]
table[row sep=crcr] {%
x	y\\
14	-2.8\\
14	-0.6\\
18	-0.6\\
18	-2.8\\
}--cycle;

\addplot[area legend, line width=1.5pt, draw=black, fill=gray, forget plot, thick]
table[row sep=crcr] {%
x	y\\
16.5175	-0.7\\
18	-1.7\\
16.5175	-2.7\\
}--cycle;
\end{axis}
\draw[dashed, white, ultra thick, dash pattern=on 12pt off 12pt] (2.75,5.22) -- (10,5.22);
\draw[dashed, white, ultra thick, dash pattern=on 12pt off 12pt] (2.75,4.46) -- (10,4.46);
\end{tikzpicture}%

%% file: 05_framework.tex
Section~\ref{sec:challenges} and \ref{sec:architecture} described the basic relation between decision making under uncertainty and architectural decisions for trajectory generation.
To further illustrate this example, we present an exemplary implementation of an MPC-based optimization framework (based on~\cite{nolte2017}), which supports the inclusion of uncertainty emerging from occluded parts of the environment as well as the selection of soft constraints to support different driving strategies.
The planner is implemented in the ACADO Optimization Framework~\cite{houska2011}, using the QPOASES solver~\cite{ferreau2014} and the provided MATLAB/Simulink interface.
A linear discrete time single track model similar to \cite{nolte2017} is used with rear steering fixed at \SI{0}{\radian}.

A slack term $\varepsilon$ is modeled as an additional system input with linear constraints
\begin{equation}
	\mathbf{u}_{\mathrm{soft-constraint}}(k) = \begin{bmatrix}
		u_{\mathrm{vehicle}} \\
		\varepsilon
	\end{bmatrix}(k), \quad
	\varepsilon \geq 0.
\end{equation}
In this way, the slack term can be weighted in the cost-function and can be subject to additional constraints itself.
Constraining the slack term permits controlling how far soft constraints may be violated.

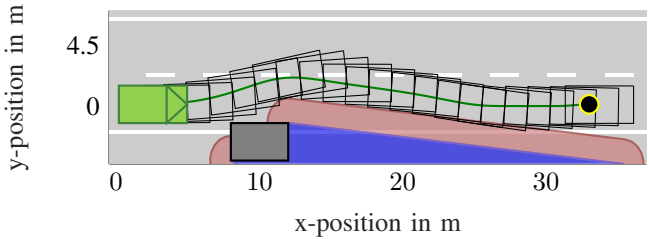
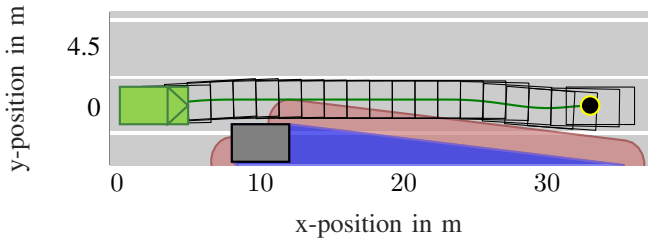
\begin{figure}[bthp]
	\centering
	\begingroup
	\captionsetup[subfigure]{width=\columnwidth}
	\subfloat[Exemplary parameterization with no constraint on the lane center line and soft constraints on the areas adjacent to occluded space. Vehicle avoids area where objects can emerge from occlusion. Consideration of uncertainty is prioritzed compared to staying inside the ego lane.]{\input{./figures/tikz/overtake}
		\label{fig:overtaking}}
	\endgroup
	\hfil
	\vspace{0.75em}
	\begingroup
	\captionsetup[subfigure]{width=\columnwidth}
	\subfloat[Lane center line modeled as hard constraint, vehicle stays at a leftmost position inside the ego lane. Lane keeping is prioritized over a safety distance to the occluded areas.]{\input{./figures/tikz/hard_centerline}
		\label{fig:hardCenterline}}
	\endgroup	
	\caption{Figures a) and b) show a comparison of a different set of constraints with similar optimization parameters. Green: ego vehicle, red: possible occupancy generated by an object emerging from the occluded space in blue.}
	\label{fig_sim}
	\vspace{-0.5em}
\end{figure}

The relevant constraints are put on the lateral deviation $e(k)$ from a reference path, resulting in lower and upper soft constraints for the lateral path deviation:
\begin{eqnarray}
	e(k) &\geq& e_{\mathrm{slack, min}}(k) - \varepsilon, \\
	e(k) &\leq& e_{\mathrm{slack, max}}(k) + \varepsilon.
\end{eqnarray}
The application of these slack constraints shall be illustrated in the following example:
A static object is parked at the edge of the ego lane, creating an occluded area which extends over parts of the ego lane.
Solid lines are modeled as hard constraints, the dashed center line in Figure~\ref{fig:overtaking} is not assumed to be relevant in this case, as no object on the opposite lane is detected.

Possible objects emerging from the occluded area (blue) are modeled by an occupancy (red) according to the assumptions made in Section~\ref{sec:challenges}.
The possible occupancy created by those assumptions is modeled as soft constraints on the lower bound of the lateral deviation.
In the target pose (depicted as a black dot), orientation (\SI{0}{\radian}) and x-position (\SI{0}{\meter}) are imposed as hard constraints.

The semantics in the given scenario differ from the overtaking example in Section \ref{sec:architecture}:
As the obstacle is only blocking parts of the ego lane, the situation can be resolved by performing a passing maneuver which does not require a lane change.
This passing maneuver can be executed inside the ego lane, if required.
Figure~\ref{fig:overtaking} and \ref{fig:hardCenterline} show two configurations of the MPC-Problem, including a dashed and a solid lane center line.

The explicit decisions modeled in both situations are the following:
Safety margins are prioritized  over staying in the ego lane by not modeling the dashed center line as a constraint in Figure~\ref{fig:overtaking}, as dashed centerlines may be passed and no oncoming traffic is visible.
Uncertainty about emerging objects is taken into account by avoiding the modeled occupancies.

By imposing a hard upper constraint on the lateral path deviation at the position of the solid lane center line in Figure~\ref{fig:hardCenterline}, traffic rules gain explicit precedence over accounting for uncertainty.
The vehicle is still able to perform an evasion maneuver, while staying in its lane and keeping to the left most lateral position, without violating the hard constraint.
The latter strategy could be chosen e.g. if available semantic information (time of day, location, etc.) makes emerging pedestrians unlikely.

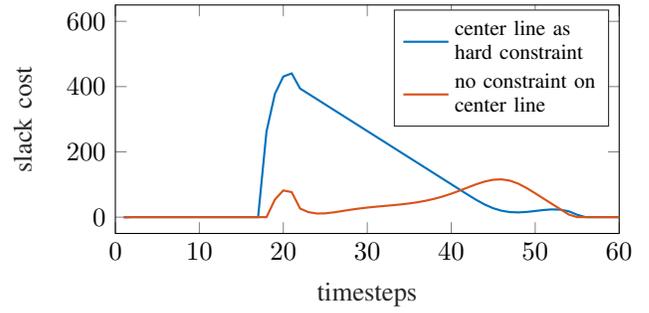
\begin{figure}[bt]
	\input{figures/tikz/slack_cost}
	\caption{Illustration of cost caused by the slack term}
	\label{fig:slack}
		\vspace{-1em}
\end{figure}

The slack value in the example is chosen to be greater than the highest weight on inputs and states in the cost function by a factor $3$.
The resulting slack values are depicted in Figure~\ref{fig:slack}.
It is visible how the second scenario incurs higher cost for the slack variable.
It is also visible how the vehicle still violates the soft constraints slightly in the first scenario.
This is due to two facts: On the one hand, a factor $3$ compared to the weights e.g. on the yaw rate means that the slack term does not dominate the cost function entirely, which is desired.
On the other hand, the optimization algorithm must still respect the yaw angle constraint in the target pose, which causes the slight curve near the end of the green trajectory.

The MPC framework (code generated from Simulink and compiled with MSVC 10.0) for trajectory planning runs with an average execution time of ca. \SI{95}{\milli\second} on an Intel i7-4800MQ CPU (\SI{2.7}{\giga\hertz}).

%% file: figures/tikz/overtake.tex
%
%
\definecolor{mycolor1}{rgb}{0.80000,0.50000,0.50000}%
\definecolor{mycolor2}{rgb}{0.00000,0.44700,0.74100}%
\definecolor{mycolor3}{rgb}{1.00000,1.00000,0.00000}%
\definecolor{mycolor4}{rgb}{0.85000,0.32500,0.09800}%
\definecolor{mycolor5}{rgb}{0.25098,0.50196,0.25098}%
\definecolor{mycolor6}{rgb}{0.57255,0.81569,0.31373}%
\begin{tikzpicture}

\begin{axis}[%
width=2.93in*0.96,
height=0.704in*1.14,
at={(1.083in,1.381in)},
scale only axis,
xmin=-0.5,
xmax=37,
xlabel style={font=\color{white!15!black}},
xlabel={x-position in m},
ymin=-5,
ymax=4,
ylabel style={font=\color{white!15!black}},
ylabel={y-position in m},
axis background/.style={fill=white},
ytick={-5.2, -1.6, 2},
yticklabels={-4.5,0, 4.5},
axis x line*=bottom,
axis y line*=left
]

\addplot[area legend, line width=1.5pt, draw=white!80!black, fill=white!80!black, forget plot, thin]
table[row sep=crcr] {%
x	y\\
-0.5	-5\\
37	-5\\
37	4\\
-0.5	4\\
}--cycle;

\addplot[area legend, draw=none, fill=white, forget plot]
table[row sep=crcr] {%
x	y\\
2.1	0.1\\
2.1	0.3\\
4.1	0.3\\
4.1	0.1\\
}--cycle;

\addplot[area legend, draw=none, fill=white, forget plot]
table[row sep=crcr] {%
x	y\\
6.1	0.1\\
6.1	0.3\\
8.1	0.3\\
8.1	0.1\\
}--cycle;

\addplot[area legend, draw=none, fill=white, forget plot]
table[row sep=crcr] {%
x	y\\
10.1	0.1\\
10.1	0.3\\
12.1	0.3\\
12.1	0.1\\
}--cycle;

\addplot[area legend, draw=none, fill=white, forget plot]
table[row sep=crcr] {%
x	y\\
14.1	0.1\\
14.1	0.3\\
16.1	0.3\\
16.1	0.1\\
}--cycle;

\addplot[area legend, draw=none, fill=white, forget plot]
table[row sep=crcr] {%
x	y\\
18.1	0.1\\
18.1	0.3\\
20.1	0.3\\
20.1	0.1\\
}--cycle;

\addplot[area legend, draw=none, fill=white, forget plot]
table[row sep=crcr] {%
x	y\\
22.1	0.1\\
22.1	0.3\\
24.1	0.3\\
24.1	0.1\\
}--cycle;

\addplot[area legend, draw=none, fill=white, forget plot]
table[row sep=crcr] {%
x	y\\
26.1	0.1\\
26.1	0.3\\
28.1	0.3\\
28.1	0.1\\
}--cycle;

\addplot[area legend, draw=none, fill=white, forget plot]
table[row sep=crcr] {%
x	y\\
30.1	0.1\\
30.1	0.3\\
32.1	0.3\\
32.1	0.1\\
}--cycle;

\addplot[area legend, draw=none, fill=white, forget plot]
table[row sep=crcr] {%
x	y\\
34.1	0.1\\
34.1	0.3\\
36.1	0.3\\
36.1	0.1\\
}--cycle;

\addplot[area legend, line width=0.0pt, draw=none, fill=white, forget plot]
table[row sep=crcr] {%
x	y\\
-0.5	-3\\
37	-3\\
37	-3.2\\
-0.5	-3.2\\
}--cycle;

\addplot[area legend, line width=0.0pt, draw=none, fill=white, forget plot]
table[row sep=crcr] {%
x	y\\
-0.5	3.4\\
37	3.4\\
37	3.6\\
-0.5	3.6\\
}--cycle;

\addplot[area legend, line width=1.5pt, draw=red!60!teal, fill=mycolor1, forget plot, thick, opacity=0.7]
table[row sep=crcr] {%
x	y\\
8.4061	-6.45\\
8.3324039043917	-6.44812599089033\\
8.25889830141032	-6.4425088075753\\
8.18577319128979	-6.43316296957589\\
8.11321759075081	-6.4201126343894\\
8.04141904442286	-6.40339153504621\\
7.9705631400713	-6.38304289291527\\
7.90083302888263	-6.3591193059838\\
7.83240895204803	-6.33168261289985\\
7.76546777486859	-6.30080373312932\\
7.35936777486859	-6.10080373312932\\
7.00544447750029	-5.8551584301257\\
6.73931799451183	-5.51636644326652\\
6.5844812902475	-5.11433546147487\\
6.55460295240899	-4.68455575018571\\
6.65232056413955	-4.264967161607\\
6.86900786493925	-3.89260990173426\\
7.1855362548492	-3.6003547158285\\
7.57396341826592	-3.41400114320958\\
8	-3.35\\
8	-4.8\\
8.4061	-5\\
}--cycle;

\addplot[area legend, line width=1.5pt, draw=red!60!teal, fill=mycolor1, forget plot, thick, opacity=0.7]
table[row sep=crcr] {%
x	y\\
10.55	-2.6\\
10.5749872326207	-2.33197273868853\\
10.6490877418016	-2.07318307672476\\
10.7697476424939	-1.83255023822366\\
10.9328083733092	-1.61836766968427\\
11.132650021955	-1.43801720512509\\
11.3623850163456	-1.29771465007879\\
11.614095505797	-1.20229555293191\\
11.8791062509256	-1.15504854703186\\
12.1482836170619	-1.15760200745043\\
35.4937836170619	-3.55760200745043\\
35.9632786686269	-3.68818845995717\\
36.3629953110419	-3.96694468105155\\
36.6477853499212	-4.3623850163456\\
36.7854816081446	-4.8298442824786\\
36.7605312245781	-5.31652272188415\\
36.5757523575062	-5.76744976177634\\
36.252015872246	-6.13169296779913\\
35.8258879664871	-6.36811088792333\\
35.3455	-6.45\\
35.3455	-5\\
12	-2.6\\
}--cycle;

\addplot [color=mycolor3, draw=none, mark size=3.5pt, mark=*, mark options={solid, fill=black}, forget plot, thick]
  table[row sep=crcr]{%
33	-1.5\\
};
\addplot [color=black!50!green, line width=2.0pt, forget plot, thick]
  table[row sep=crcr]{%
1.3	-1.5\\
1.828	-1.49947325097241\\
2.356	-1.49595949044327\\
2.884	-1.48699557143687\\
3.412	-1.47057532446344\\
3.94	-1.4449946649714\\
4.468	-1.40875223272948\\
4.996	-1.36049362183895\\
5.524	-1.29897224854248\\
6.052	-1.22302319954177\\
6.58	-1.13172673991839\\
7.108	-1.02471175592266\\
7.636	-0.902498196589236\\
8.164	-0.766856128944671\\
8.692	-0.621158955775766\\
9.22	-0.470673648247743\\
9.748	-0.322679026996517\\
10.276	-0.186226112528428\\
10.804	-0.0712417870876942\\
11.332	0.0133703069364055\\
11.86	0.0623431265173635\\
12.388	0.0757169532903608\\
12.916	0.0583573020914638\\
13.444	0.0177809559331223\\
13.972	-0.0380811699679295\\
14.5	-0.102374826458288\\
15.028	-0.170020220056314\\
15.556	-0.237807967299393\\
16.084	-0.304129513519773\\
16.612	-0.368550808697704\\
17.14	-0.431379333157934\\
17.668	-0.493311342650563\\
18.196	-0.555192752348519\\
18.724	-0.617888700066938\\
19.252	-0.682233293537306\\
19.78	-0.749019586296263\\
20.308	-0.818987334318825\\
20.836	-0.892770462313737\\
21.364	-0.970776854332631\\
21.892	-1.05299083831835\\
22.42	-1.13871473769985\\
22.948	-1.22630018679932\\
23.476	-1.3129594563427\\
24.004	-1.39478315041685\\
24.532	-1.46710669805959\\
25.06	-1.52533764350655\\
25.588	-1.56624194776527\\
26.116	-1.589444098364\\
26.644	-1.59847012306829\\
27.172	-1.6\\
27.7	-1.6\\
28.228	-1.6\\
28.756	-1.6\\
29.284	-1.59998059333701\\
29.812	-1.59941218154677\\
30.34	-1.6\\
30.868	-1.59862106096974\\
31.396	-1.58234491494359\\
31.924	-1.54954787593347\\
32.452	-1.51671185886226\\
};
\addplot [color=black, forget plot]
  table[row sep=crcr]{%
4.947	-0.4\\
4.947	-2.6\\
0.182	-2.6\\
0.182	-0.4\\
4.947	-0.4\\
};
\addplot [color=black, forget plot]
  table[row sep=crcr]{%
6.52496161311055	-0.36717079413668\\
6.5369304432584	-2.56713823641961\\
1.77200096031378	-2.59306163444438\\
1.76003213016593	-0.393094192161447\\
6.52496161311055	-0.36717079413668\\
};
\addplot [color=black, forget plot]
  table[row sep=crcr]{%
8.08498581340336	-0.213753226186435\\
8.14251953653388	-2.41300079540249\\
3.3791492332091	-2.53761360936475\\
3.32161551007858	-0.338366040148691\\
8.08498581340336	-0.213753226186435\\
};
\addplot [color=black, forget plot]
  table[row sep=crcr]{%
9.62061213018995	0.109499394736161\\
9.76225473305175	-2.08593618233153\\
5.00714085817559	-2.39272118352993\\
4.86549825531379	-0.19728560646224\\
9.62061213018995	0.109499394736161\\
};
\addplot [color=black, forget plot]
  table[row sep=crcr]{%
11.1273954320503	0.620908163203858\\
11.3874618180719	-1.56366627472717\\
6.65587218318947	-2.12694642445117\\
6.3958057971679	0.0576280134798624\\
11.1273954320503	0.620908163203858\\
};
\addplot [color=black, forget plot]
  table[row sep=crcr]{%
12.6248221013549	1.23748192374363\\
13.0014807192693	-0.930034693390089\\
8.3068367735229	-1.74584301810021\\
7.93017815560847	0.42167359903351\\
12.6248221013549	1.23748192374363\\
};
\addplot [color=black, forget plot]
  table[row sep=crcr]{%
14.1846836952443	1.68420310031536\\
14.5915158252464	-0.477853239838129\\
9.90869834305036	-1.35901464868372\\
9.5018662130482	0.803041691469771\\
14.1846836952443	1.68420310031536\\
};
\addplot [color=black, forget plot]
  table[row sep=crcr]{%
15.8573146480197	1.64876014183799\\
16.1485094056474	-0.531883252467953\\
11.4254340538893	-1.16258462523887\\
11.1342392962616	1.01805876906707\\
15.8573146480197	1.64876014183799\\
};
\addplot [color=black, forget plot]
  table[row sep=crcr]{%
17.5523010688746	1.26270830424194\\
17.6744479533161	-0.9338982009431\\
12.9167979545858	-1.19845724838118\\
12.7946510701443	0.998149256803862\\
17.5523010688746	1.26270830424194\\
};
\addplot [color=black, forget plot]
  table[row sep=crcr]{%
19.2054561537011	0.854015636682942\\
19.2005255275803	-1.34597883806604\\
14.4355374947717	-1.33529955012702\\
14.4404681208925	0.864694924621969\\
19.2054561537011	0.854015636682942\\
};
\addplot [color=black, forget plot]
  table[row sep=crcr]{%
20.8264514193127	0.528143951107605\\
20.7421976629224	-1.67024211701922\\
15.9806932926386	-1.48775614011029\\
16.0649470490289	0.710629928016536\\
20.8264514193127	0.528143951107605\\
};
\addplot [color=black, forget plot]
  table[row sep=crcr]{%
22.4333247083373	0.249013375209561\\
22.2940429122801	-1.94657323918011\\
17.5386019042952	-1.64490153090183\\
17.6778837003524	0.550685083487838\\
22.4333247083373	0.249013375209561\\
};
\addplot [color=black, forget plot]
  table[row sep=crcr]{%
24.0354659229205	-0.0337441914220085\\
23.8480066852862	-2.22574305871769\\
19.100336411348	-1.8197233917507\\
19.2877956489823	0.37227547554499\\
24.0354659229205	-0.0337441914220085\\
};
\addplot [color=black, forget plot]
  table[row sep=crcr]{%
25.6362957127469	-0.35237780996011\\
25.399234880006	-2.53956828213673\\
20.6619791527689	-2.02611606940468\\
20.8990399855098	0.161074402771941\\
25.6362957127469	-0.35237780996011\\
};
\addplot [color=black, forget plot]
  table[row sep=crcr]{%
27.2323735809191	-0.680293256381516\\
26.9557283982768	-2.8628301845733\\
22.2285518242614	-2.2636418685322\\
22.5051970069037	-0.0811049403404158\\
27.2323735809191	-0.680293256381516\\
};
\addplot [color=black, forget plot]
  table[row sep=crcr]{%
28.8129492216699	-0.872665337087093\\
28.5479371788869	-3.05664533177111\\
23.8176350540373	-2.48265333910702\\
24.0826470968202	-0.298673344423003\\
28.8129492216699	-0.872665337087093\\
};
\addplot [color=black, forget plot]
  table[row sep=crcr]{%
30.3727163105743	-0.819186044405549\\
30.1817535746307	-3.01088247262752\\
25.4347383562318	-2.59727454682251\\
25.6257010921753	-0.405578118600535\\
30.3727163105743	-0.819186044405549\\
};
\addplot [color=black, forget plot]
  table[row sep=crcr]{%
31.9339198402774	-0.718656061779336\\
31.803191277503	-2.91476854589187\\
27.0466112835047	-2.63162236333726\\
27.1773398462791	-0.435509879224728\\
31.9339198402774	-0.718656061779336\\
};
\addplot [color=black, forget plot]
  table[row sep=crcr]{%
33.5010081040053	-0.649765821798697\\
33.4108666765235	-2.84791834677315\\
28.6498681394766	-2.65268020952279\\
28.7400095669584	-0.454527684548342\\
33.5010081040053	-0.649765821798697\\
};
\addplot [color=black, forget plot]
  table[row sep=crcr]{%
35.0631970980318	-0.551677924986771\\
35.0214922012892	-2.75128259437345\\
30.2573484514585	-2.66095357938318\\
30.2990533482011	-0.461348909996502\\
35.0631970980318	-0.551677924986771\\
};
\addplot [color=black, forget plot]
  table[row sep=crcr]{%
36.0991631901089	-0.417253053488096\\
36.0988367295844	-2.61725302926617\\
31.3338367820469	-2.61654594544842\\
31.3341632425713	-0.416545969670342\\
36.0991631901089	-0.417253053488096\\
};

\addplot[area legend, line width=1.5pt, draw=white!40!blue, fill=blue!80!gray, forget plot, thick, opacity=0.7]
table[row sep=crcr] {%
x	y\\
8.4061	-5\\
8	-4.8\\
8	-2.6\\
12	-2.6\\
35.3455	-5\\
}--cycle;

\addplot[area legend, line width=1.5pt, draw=mycolor5, fill=mycolor6, forget plot, thick]
table[row sep=crcr] {%
x	y\\
0.182	-2.6\\
0.182	-0.4\\
4.947	-0.4\\
4.947	-2.6\\
}--cycle;

\addplot[area legend, line width=1.5pt, draw=mycolor5, fill=mycolor6, forget plot, thick]
table[row sep=crcr] {%
x	y\\
3.5175	-0.4\\
4.947	-1.5\\
3.5175	-2.6\\
}--cycle;

\addplot[area legend, line width=1.5pt, draw=black, fill=gray, forget plot, thick] 
table[row sep=crcr] {%
x	y\\
8	-4.8\\
8	-2.6\\
12	-2.6\\
12	-4.8\\
}--cycle;
\end{axis}
\end{tikzpicture}%

%% file: figures/tikz/hard_centerline.tex
%
%
\definecolor{mycolor1}{rgb}{0.80000,0.50000,0.50000}%
\definecolor{mycolor2}{rgb}{0.00000,0.44700,0.74100}%
\definecolor{mycolor3}{rgb}{1.00000,1.00000,0.00000}%
\definecolor{mycolor4}{rgb}{0.85000,0.32500,0.09800}%
\definecolor{mycolor5}{rgb}{0.25098,0.50196,0.25098}%
\definecolor{mycolor6}{rgb}{0.57255,0.81569,0.31373}%
\begin{tikzpicture}

\begin{axis}[%
width=2.93in*0.96,
height=0.704in*1.14,
at={(1.083in,1.381in)},
scale only axis,
xmin=-0.5,
xmax=37,
xlabel style={font=\color{white!15!black}},
xlabel={x-position in m},
ymin=-5,
ymax=4,
ytick={-5.2, -1.6, 2},
yticklabels={-4.5,0, 4.5},
ylabel style={font=\color{white!15!black}},
ylabel={y-position in m},
axis background/.style={fill=white},
axis x line*=bottom,
axis y line*=left
]

\addplot[area legend, line width=1.5pt, draw=white!80!black, fill=white!80!black, forget plot, thick]
table[row sep=crcr] {%
x	y\\
-0.5	-5\\
37	-5\\
37	4\\
-0.5	4\\
}--cycle;

\addplot[area legend, draw=white, fill=white, forget plot, thin]
table[row sep=crcr] {%
x	y\\
-0.5	0.1\\
-0.5	0.2\\
37	0.2\\
37	0.1\\
}--cycle;

\addplot[area legend, line width=0.0pt, draw=none, fill=white, forget plot, thin]
table[row sep=crcr] {%
x	y\\
-0.5	-3\\
37	-3\\
37	-3.2\\
-0.5	-3.2\\
}--cycle;

\addplot[area legend, line width=0.0pt, draw=none, fill=white, forget plot, thin]
table[row sep=crcr] {%
x	y\\
-0.5	3.4\\
37	3.4\\
37	3.6\\
-0.5	3.6\\
}--cycle;

\addplot[area legend, line width=1.5pt, draw=black, fill=gray, forget plot, thick]
table[row sep=crcr] {%
x	y\\
8	-4.8\\
8	-2.6\\
12	-2.6\\
12	-4.8\\
}--cycle;

\addplot[area legend, line width=1.5pt, draw=red!60!teal, fill=mycolor1, forget plot, , thick,opacity=0.7]
table[row sep=crcr] {%
x	y\\
8.4061	-6.45\\
8.3324039043917	-6.44812599089033\\
8.25889830141032	-6.4425088075753\\
8.18577319128979	-6.43316296957589\\
8.11321759075081	-6.4201126343894\\
8.04141904442286	-6.40339153504621\\
7.9705631400713	-6.38304289291527\\
7.90083302888263	-6.3591193059838\\
7.83240895204803	-6.33168261289985\\
7.76546777486859	-6.30080373312932\\
7.35936777486859	-6.10080373312932\\
7.00544447750029	-5.8551584301257\\
6.73931799451183	-5.51636644326652\\
6.5844812902475	-5.11433546147487\\
6.55460295240899	-4.68455575018571\\
6.65232056413955	-4.264967161607\\
6.86900786493925	-3.89260990173426\\
7.1855362548492	-3.6003547158285\\
7.57396341826592	-3.41400114320958\\
8	-3.35\\
8	-4.8\\
8.4061	-5\\
}--cycle;

\addplot[area legend, line width=1.5pt, draw=red!60!teal, fill=mycolor1, forget plot,, thick, opacity=0.7]
table[row sep=crcr] {%
x	y\\
10.55	-2.6\\
10.5749872326207	-2.33197273868853\\
10.6490877418016	-2.07318307672476\\
10.7697476424939	-1.83255023822366\\
10.9328083733092	-1.61836766968427\\
11.132650021955	-1.43801720512509\\
11.3623850163456	-1.29771465007879\\
11.614095505797	-1.20229555293191\\
11.8791062509256	-1.15504854703186\\
12.1482836170619	-1.15760200745043\\
35.4937836170619	-3.55760200745043\\
35.9632786686269	-3.68818845995717\\
36.3629953110419	-3.96694468105155\\
36.6477853499212	-4.3623850163456\\
36.7854816081446	-4.8298442824786\\
36.7605312245781	-5.31652272188415\\
36.5757523575062	-5.76744976177634\\
36.252015872246	-6.13169296779913\\
35.8258879664871	-6.36811088792333\\
35.3455	-6.45\\
35.3455	-5\\
12	-2.6\\
}--cycle;

\addplot [color=mycolor3, draw=none, mark size=3.5pt, mark=*, mark options={solid, fill=black}, forget plot, , thick]
  table[row sep=crcr]{%
33	-1.5\\
};
\addplot [color=black!50!green, line width=2.0pt, forget plot, thick]
  table[row sep=crcr]{%
1.3	-1.5\\
1.828	-1.49773866554535\\
2.356	-1.48400341501693\\
2.884	-1.45542314631283\\
3.412	-1.41446103835645\\
3.94	-1.36605725020254\\
4.468	-1.31565969978774\\
4.996	-1.26813541306614\\
5.524	-1.22720115868292\\
6.052	-1.19514199823136\\
6.58	-1.1726981362255\\
7.108	-1.15910314119489\\
7.636	-1.15236353661284\\
8.164	-1.15\\
8.692	-1.15\\
9.22	-1.15053145832181\\
9.748	-1.15036288254445\\
10.276	-1.15\\
10.804	-1.15\\
11.332	-1.15\\
11.86	-1.15\\
12.388	-1.15\\
12.916	-1.15\\
13.444	-1.15\\
13.972	-1.15\\
14.5	-1.15\\
15.028	-1.15\\
15.556	-1.15\\
16.084	-1.15\\
16.612	-1.15\\
17.14	-1.15\\
17.668	-1.15\\
18.196	-1.15\\
18.724	-1.15\\
19.252	-1.15\\
19.78	-1.15\\
20.308	-1.15\\
20.836	-1.15\\
21.364	-1.15\\
21.892	-1.1500417093735\\
22.42	-1.15\\
22.948	-1.15\\
23.476	-1.15230991114518\\
24.004	-1.16094955909615\\
24.532	-1.1794455051625\\
25.06	-1.20949452005816\\
25.588	-1.25089206988789\\
26.116	-1.3020053905099\\
26.644	-1.3602753265008\\
27.172	-1.42252662092331\\
27.7	-1.48506605156847\\
28.228	-1.54367423024061\\
28.756	-1.59366665455574\\
29.284	-1.63021746519317\\
29.812	-1.64909439093044\\
30.34	-1.64781792219483\\
30.868	-1.62701149899547\\
31.396	-1.59145831154583\\
31.924	-1.55032693386418\\
32.452	-1.51594355065645\\
};
\addplot [color=black, forget plot, , thin]
  table[row sep=crcr]{%
4.947	-0.4\\
4.947	-2.6\\
0.182	-2.6\\
0.182	-0.4\\
4.947	-0.4\\
};
\addplot [color=black, forget plot, , thin]
  table[row sep=crcr]{%
6.51159484079375	-0.29302705588299\\
6.54933200794048	-2.49270337439345\\
1.78503307262123	-2.57443864778172\\
1.7472959054745	-0.374762329271255\\
6.51159484079375	-0.29302705588299\\
};
\addplot [color=black, forget plot, thin]
  table[row sep=crcr]{%
8.06451444179506	-0.0603936443398945\\
8.15878593103002	-2.25837291774477\\
3.39816264113264	-2.46255639329231\\
3.30389115189768	-0.264577119887431\\
8.06451444179506	-0.0603936443398945\\
};
\addplot [color=black, forget plot, thin]
  table[row sep=crcr]{%
9.6419781474957	0.0786970420336617\\
9.74760934731115	-2.11876559692645\\
4.98810504065436	-2.34755317289036\\
4.88247384083891	-0.150090533930248\\
9.6419781474957	0.0786970420336617\\
};
\addplot [color=black, forget plot, thin]
  table[row sep=crcr]{%
11.2409420904106	0.0784904452980599\\
11.3203098123078	-2.12007744419748\\
6.55841163346857	-2.29198071457924\\
6.47904391157143	-0.0934128250837074\\
11.2409420904106	0.0784904452980599\\
};
\addplot [color=black, forget plot, thin]
  table[row sep=crcr]{%
12.8392598489221	0.0375624546247466\\
12.8925962641438	-2.16179091098327\\
8.12899681545188	-2.27731273758849\\
8.07566040023016	-0.0779593719804716\\
12.8392598489221	0.0375624546247466\\
};
\addplot [color=black, forget plot, thin]
  table[row sep=crcr]{%
14.4318430959362	0.0116210532130443\\
14.4691103165046	-2.18806327738331\\
9.70479402773566	-2.26878068920517\\
9.66752680716733	-0.0690963586088236\\
14.4318430959362	0.0116210532130443\\
};
\addplot [color=black, forget plot, thin]
  table[row sep=crcr]{%
16.0218217879374	-0.00721908770244761\\
16.0476745722636	-2.20706718099039\\
11.2830035883923	-2.26306196158795\\
11.257150804066	-0.0632138683000127\\
16.0218217879374	-0.00721908770244761\\
};
\addplot [color=black, forget plot, thin]
  table[row sep=crcr]{%
17.6099211630141	-0.0203369477166548\\
17.627836973384	-2.22026399735689\\
12.8629949772314	-2.2590680139081\\
12.8450791668615	-0.059140964267868\\
17.6099211630141	-0.0203369477166548\\
};
\addplot [color=black, forget plot, thin]
  table[row sep=crcr]{%
19.1967339946758	-0.0294354190800739\\
19.2091498478931	-2.22940038393494\\
14.4442257308324	-2.25629199328968\\
14.4318098776151	-0.0563270284348105\\
19.1967339946758	-0.0294354190800739\\
};
\addplot [color=black, forget plot, thin]
  table[row sep=crcr]{%
20.7826699628532	-0.0357448489114052\\
20.7912742515594	-2.23572802298703\\
16.026310694982	-2.25436413011672\\
16.0177064062757	-0.0543809560410957\\
20.7826699628532	-0.0357448489114052\\
};
\addplot [color=black, forget plot, thin]
  table[row sep=crcr]{%
22.3680051804639	-0.0401192655278837\\
22.3739680279392	-2.24011118470622\\
17.6089855302643	-2.25302617026072\\
17.603022682789	-0.0530342510823809\\
22.3680051804639	-0.0401192655278837\\
};
\addplot [color=black, forget plot, thin]
  table[row sep=crcr]{%
23.9529277191348	-0.0431527104957683\\
23.9570594176942	-2.24314883073485\\
19.1920678209037	-2.25209771420566\\
19.1879361223442	-0.0521015939665812\\
23.9529277191348	-0.0431527104957683\\
};
\addplot [color=black, forget plot, thin]
  table[row sep=crcr]{%
25.5375875777507	-0.0453697159971986\\
25.5404064348709	-2.24536791009749\\
20.7754103462855	-2.25147329836021\\
20.7725914891653	-0.0514751042599197\\
25.5375875777507	-0.0453697159971986\\
};
\addplot [color=black, forget plot, thin]
  table[row sep=crcr]{%
27.1236622997355	-0.0545081316382809\\
27.1223363755331	-2.25450773207575\\
22.3573372409492	-2.25163590079199\\
22.3586631651516	-0.0516363003545199\\
27.1236622997355	-0.0545081316382809\\
};
\addplot [color=black, forget plot, thin]
  table[row sep=crcr]{%
28.727267782546	-0.17907906447706\\
28.6854120156755	-2.37868086816328\\
23.9212744726915	-2.28802508219167\\
23.9631302395619	-0.0884232785054426\\
28.727267782546	-0.17907906447706\\
};
\addplot [color=black, forget plot, thin]
  table[row sep=crcr]{%
30.3403199164059	-0.439499525569204\\
30.2329955693362	-2.63688012176233\\
25.4736689598543	-2.40442534276834\\
25.5809933069239	-0.207044746575211\\
30.3403199164059	-0.439499525569204\\
};
\addplot [color=black, forget plot, thin]
  table[row sep=crcr]{%
31.9417839846534	-0.696079093791654\\
31.7910830440466	-2.89091148522847\\
27.0372756144119	-2.56450694795977\\
27.1879765550186	-0.369674556522952\\
31.9417839846534	-0.696079093791654\\
};
\addplot [color=black, forget plot, thin]
  table[row sep=crcr]{%
33.5159747728646	-0.759611461368497\\
33.3900709177415	-2.9560058292508\\
28.6328803891236	-2.68330952485899\\
28.7587842442468	-0.486915156976685\\
33.5159747728646	-0.759611461368497\\
};
\addplot [color=black, forget plot, thin]
  table[row sep=crcr]{%
35.0605635742268	-0.551466575785282\\
35.024453845343	-2.75117021206903\\
30.2600957422103	-2.67295982200953\\
30.296205471094	-0.473256185725777\\
35.0605635742268	-0.551466575785282\\
};
\addplot [color=black, forget plot, thin]
  table[row sep=crcr]{%
36.0988136223526	-0.415325813281867\\
36.0991862730082	-2.61532578172084\\
31.3341863413665	-2.61613290916368\\
31.3338136907109	-0.416132940724706\\
36.0988136223526	-0.415325813281867\\
};

\addplot[area legend, line width=1.5pt, draw=white!40!blue, fill=blue!80!gray, forget plot, thick,opacity=0.7]
table[row sep=crcr] {%
x	y\\
8.4061	-5\\
8	-4.8\\
8	-2.6\\
12	-2.6\\
35.3455	-5\\
}--cycle;

\addplot[area legend, line width=1.5pt, draw=mycolor5, fill=mycolor6, forget plot, thick]
table[row sep=crcr] {%
x	y\\
0.182	-2.6\\
0.182	-0.4\\
4.947	-0.4\\
4.947	-2.6\\
}--cycle;

\addplot[area legend, line width=1.5pt, draw=mycolor5, fill=mycolor6, forget plot, thick]
table[row sep=crcr] {%
x	y\\
3.5175	-0.4\\
4.947	-1.5\\
3.5175	-2.6\\
}--cycle;
\addplot[area legend, line width=1.5pt, draw=black, fill=gray, forget plot, thick]
table[row sep=crcr] {%
x	y\\
8	-4.8\\
8	-2.6\\
12	-2.6\\
12	-4.8\\
}--cycle;
\end{axis}
\end{tikzpicture}%

%% file: figures/tikz/slack_cost.tex
%
%
\definecolor{mycolor1}{rgb}{0.00000,0.44700,0.74100}%
\definecolor{mycolor2}{rgb}{0.85000,0.32500,0.09800}%
\begin{tikzpicture}

\begin{axis}[%
width=2.93in*0.9,
height=0.704in*1.7,
at={(1.083in,1.381in)},
scale only axis,
xmin=0,
xmax=60,
xlabel style={font=\color{white!15!black}},
xlabel={timesteps},
ymin=-50,
ymax=650,
ylabel style={font=\color{white!15!black}},
ylabel={slack cost},
axis background/.style={fill=white},
legend style={legend cell align=left, align=left, draw=white!15!black, font=\footnotesize}
]
\addplot [color=mycolor1, thick]
  table[row sep=crcr]{%
1	-0.523598775598292\\
2	0\\
3	0\\
4	0\\
5	0\\
6	0\\
7	0\\
8	0\\
9	0\\
10	0\\
11	0\\
12	0\\
13	0\\
14	0\\
15	0\\
16	0\\
17	0\\
18	262.519287319107\\
19	377.62910868466\\
20	430.839225768317\\
21	440.391505531998\\
22	394.042209625169\\
23	377.7581291814\\
24	361.474048737633\\
25	345.189968293864\\
26	328.905887850095\\
27	312.621807406326\\
28	296.337726962557\\
29	280.053646518789\\
30	263.76956607502\\
31	247.485485631252\\
32	231.201405187483\\
33	214.917324743715\\
34	198.633244299946\\
35	182.349163856177\\
36	166.06508341241\\
37	149.781002968641\\
38	133.496922524871\\
39	117.212842081104\\
40	100.929273193625\\
41	84.6446811935657\\
42	68.3606007497967\\
43	52.5991942646894\\
44	38.526632295911\\
45	27.3090462256639\\
46	19.6101103558301\\
47	15.4956894856617\\
48	14.5438851018325\\
49	15.9907767903289\\
50	18.8332646886276\\
51	21.8746584804988\\
52	23.7410085077356\\
53	22.9200213353857\\
54	17.8802514993776\\
55	7.31287751845892\\
56	0\\
57	0\\
58	0\\
59	0\\
60	0\\
};
\addlegendentry{center line as\\ hard constraint}

\addplot [color=mycolor2, thick]
  table[row sep=crcr]{%
1	-0.523598775598286\\
2	0\\
3	0\\
4	0\\
5	0\\
6	0\\
7	0\\
8	0\\
9	0\\
10	0\\
11	0\\
12	0\\
13	0\\
14	0\\
15	0\\
16	0\\
17	0\\
18	0\\
19	54.0016448109683\\
20	81.8281336873948\\
21	76.6885675767873\\
22	26.327123638061\\
23	15.2509385539609\\
24	11.1397619576949\\
25	11.6143192842421\\
26	14.6183357875809\\
27	18.6278734232203\\
28	22.6801171523753\\
29	26.2925005747207\\
30	29.3348086843312\\
31	31.899285578632\\
32	34.1948079826514\\
33	36.4751504482699\\
34	38.9998543200266\\
35	42.0191519173686\\
36	45.770959301287\\
37	50.4772032642869\\
38	56.328061218992\\
39	63.4458983808909\\
40	71.8260131328374\\
41	81.2591025035202\\
42	91.2506567895917\\
43	100.964357208836\\
44	109.227384987313\\
45	114.640368836367\\
46	115.825572026687\\
47	111.812782860533\\
48	102.489347596385\\
49	88.9130745639034\\
50	73.0879571996486\\
51	56.8038767558801\\
52	40.5197963121116\\
53	24.2357158683421\\
54	7.94581342567826\\
55	0\\
56	0\\
57	0\\
58	0\\
59	0\\
60	0\\
};
\addlegendentry{no constraint on\\ center line}

\end{axis}
\end{tikzpicture}%

%% file: 06_conclusion.tex
In this paper we have discussed the effects of different types of uncertainty on system design for the interaction between decision making and motion planning.
We discussed preceding work, discussing motion planning as an optimization problem while interpreting the role of decision making modules for the parametrization of the optimization algorithms.
In this context we argue for a clear separation between explicit tactic decisions and implicit decisions made at the stabilization level in order to perform a trade-off between conservative and optimistic (or more risk-prone) system implementations.

We also presented an MPC framework which provides the possibility to reflect such explicit decisions in the selection of hard and soft constraints.
In the given example, we modeled epistemic uncertainty caused by occluded space in an aleatoric distribution and show how formulating soft constraints can be used to differentiate between a conservative and optimistic driving styles.
For future work, the proposed system is currently being integrated in our research vehicles \emph{MOBILE} \cite{bergmiller2014a} and \emph{Leonie} and will be validated using the environment perception framework described in \cite{rieken2015a}.

%% file: 07_acknowledgement.tex
This work has partially been funded by the Daimler and Benz Foundation.
The authors would like to thank their colleagues Gerrit Bagschik, Jens Rieken and Cordula Reisch for valuable discussions.